Exposing Hidden Alternative Conformations of Small Flexible Molecules in 2D NMR Spectroscopy Using $^{1}$H-$^{15}$N HMBC

Khodov I.A., Efimov S.V., Krestyaninov M.A., Kiselev M.G.

Abstract—Two-dimensional $^{1}$H-$^{15}$N HMBC NMR spectra of a well-known anticonvulsant—carbamazepine—dissolved in chloroform, recorded on an NMR spectrometer and obtained from quantum-chemical calculations prove the existence of hidden conformers in saturated solutions. A weaker influence of ring currents was revealed for the hidden conformation of carbamazepine dissolved in saturated solution, which provides a simple approach for discovering hidden conformations. Hidden conformers were found in three different solvents: dimethyl sulfoxide, chloroform, and dichloromethane.

Lately NMR spectroscopy has shown itself as a useful instrument for conformational analysis of small conformationally labile in solution molecules [1–11]. Nuclear Overhauser effect spectroscopy (NOESY) is the most accurate method in investigations of such kind of phenomena [12,13]. However, use of NOE-based methods is limited in cases of small highly symmetrical molecules or atomic groups, involved in conformational transformation and at the same time participating in chemical exchange, because experiments NOESY and exchange spectroscopy EXSY are actually the same pulse sequence[14].
For instance, for carbamazepine it is impossible to observe cross-peaks characterizing interproton interactions between aryl fragment (Ar) and amine group (NH$_2$) utilizing the two-dimensional NOESY approach. Molecular structure of carbamazepine belongs to the nonaxial (without rotation) symmetry point group C$_1$. On the NOESY spectrum we can obtain only cross-peaks for atoms within the aryl fragment and no cross-peaks for the amine group. Thus, changes in the molecular structure induced by the lability of the NH$_2$ group remain hidden for the NOESY registration (Figure 1).

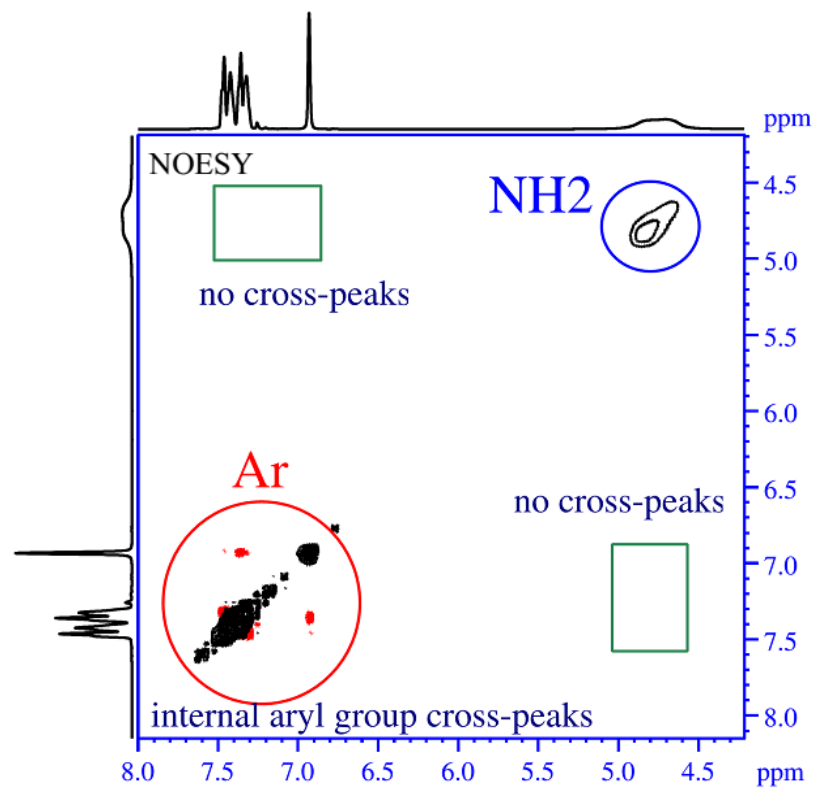

Figure 1. Two dimensional NOESY spectrum of carbamazepine in saturated solution. In this spectrum only internal aryl group cross-peaks are observed but no signals for the aryl-amine groups interaction.

Here we demonstrate a simple and generally applicable approach based on $^1$H-$^{15}$N heteronuclear multiple-bond correlation (HMBC) NMR experiments in order to establish the presence of hidden conformers of a studied drug compound. This method relies on the fact that chemical shifts of $^1$H and $^{15}$N nuclei are very sensitive to changes in the N–H bond length [15], which is an important criterion for distinguishing of carbamazepine conformers.

Carbamazepine is a drug belonging to the carboxamide group and used in the treatment of epilepsy [16]. Till now, four polymorphs of carbamazepine have been obtained and their structures were resolved, but characterization of their conformational structures still remains doubtful. Nevertheless, molecular conformation, preferable in crystal, does not always correspond to the conformation of the isolated molecule in the gas phase with the minimal energy. As it has been recently shown for carbamazepine in $CO_2$, there are solid correlations between the conformers in fluid and polymorphs in solid phase [17]. At the beginning of the characterization of the molecule in liquid phase, we decided to investigate the isolated molecule in the gas phase. Carbamazepine molecule represents a tricyclic compound consisting of two aryl rings linked by an azepine fragment with an amide group from the nitrogen atom. In general, conformational lability of the carbamazepine molecule is due to the mobility of the amine group in the amide fragment.

Thus, we optimized the geometry of the isolated carbamazepine molecule in the gas phase. *Ab initio* quantum chemical calculations were performed using the density functional theory DFT. All calculations have been performed using Gaussian 03 program package [18], B3LYP functional / 6-31G** basis set [19]; chemical shifts were calculated using the GIAO approach [20]. Two molecule conformations were determined as the minima on the potential energy surface (Figure 2). Global minimum (conf. 1) was more stable than the second local one (conf. 2) for approximately 1.75 kJ/mol. The sole difference between two geometries of these conformations is the orientation of the hydrogen atoms of the amine group, correspondingly to the plane of the amide group.

The shortest energetic interconversion pathway between these two minima was investigated with the aid of relative conformational energies *E*, obtained depending on the chosen angle between the HNH and OCN planes (Figure 2). The calculations were performed with the angle constraint between the planes for a number of values from 120° to 240° with the optimization of the amine residue at each step. Each angle value between the HNH and OCN planes corresponds to the different degree of deviation of amine group relative to the plane of the carbonyl fragment. The angle values less than 180° lead to the change of the amine hydrogen atoms direction under the plane of the amide group, and angles more than 180° lead to the inverse position that is over the plane of the amide group (Figure 2).

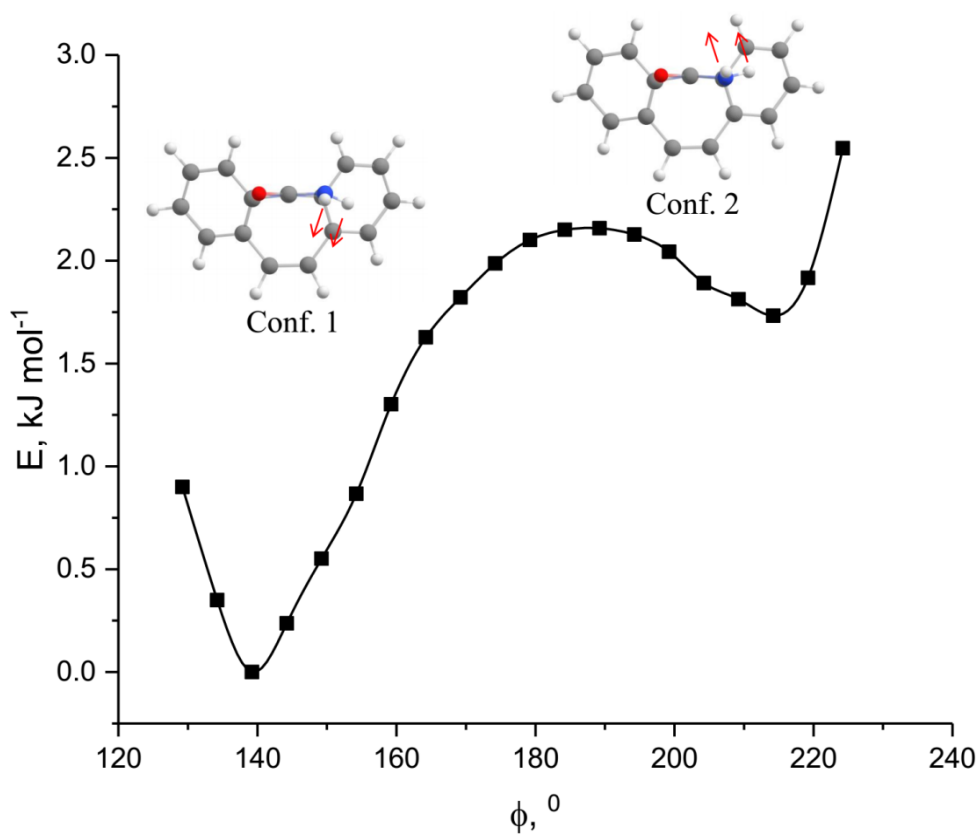

Figure 2. Conf. 1 – Conf. 2 gas-phase interconversion pathway, where *E* is the conformational energy and φ is angle between the HNH and OCN planes

Nitrogen chemical shift was determined using $^1$H-$^{15}$N HMBC experiment (see Figure 3) carried out at the natural isotope abundance. Additional signal shifted to the low-frequency region was observed in the spectra. Quantum-chemical calculations GIAO show that the chemical shift of protons in the NH$_2$ group is shifted to the high-frequency (downfield) region for the lower-energy conformer 1, whereas the signal of the higher-energy conformers 2 appears at the low frequency (high field). This is due to the fact that NH$_2$ protons in conf. 2 of carbamazepine are placed at a larger distance from the ring planes than in conf. 1, and hence the contribution to the chemical shift from the additional deshielding by ring currents is smaller in conf. 2. The influence of the ring currents for the NH$_2$ group was estimated using Eq. (1):

$$\delta_{rc} = \mu(3\cos^2\theta - 1)/R^3$$

where R is the distance of the hydrogen atom from the ring center, θ is the angle between the R vector and the ring symmetry axis, and µ is the equivalent dipole of the aromatic ring.

This contribution was found to be 0.96 ppm for conf. 1 and 0.48 ppm for conf. 2. It was also found that it is the azepine moiety which brings the largest contribution rather than the side aryl groups. The chemical shift difference is 0.48 ppm based on the quantum chemical calculation, which is in a good agreement with the experimental observations in DMSO.

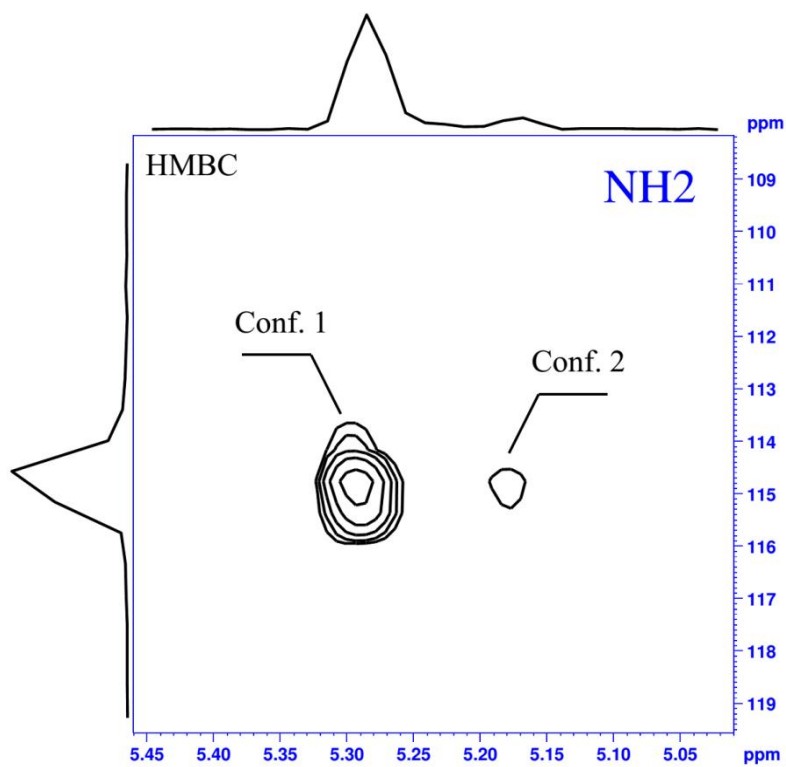

Figure 3. $^1$H-$^{15}$N HMBC spectra of carbamazepine in (a) saturated solution and (b) diluted solution in chloroform, obtained at the natural isotope abundance (300 K, 500 MHz, Bruker Avance III).

The HMBC spectra for the saturated solution of carbamazepine in deuterated chloroform (CDCl$_3$) (see Figure 3), in deuterated dichloromethane (DMC), and deuterated dimethyl sulfoxide (DMSO) were recorded with 16 scans per increment, 1024 t1 increments using the standard HMBC pulse program "hmbcgplpndqf" [21–23] available in the Bruker TOPSPIN 2.1 library. Signal of nitromethane was used as a reference for $^{15}$N chemical shift after recalculation to the liquid ammonium scale [24]; sweep width was 1000 ppm in the nitrogen dimension and 15 ppm in the proton dimension. The signal-to-noise ratio for the cross-peak corresponding to the saturated solution in CDCl$_3$ is 150:1; for the saturated solution in DMSO, 350:1; for the saturated solution of DMC, 100:1

Previous observations have shown that an additional signal can appear in experiments with the phase separation due to the presence of two different conformational polymorphs of carbamazepine in the solid phase [25]. Conformers providing different crystal packing are distinguished by rotation of the amide group. Conventional approaches face with a difficulty in finding hidden conformations since the NH$_2$ group in the carbamazepine molecule is involved in the chemical exchange. On the other hand, the method presented in this paper might be used for solving this problem without any limitations. In fact, observing two scalar N–H correlations for NH$_2$ group in the $^1$H-$^{15}$N HMBC spectra of two conformers is a strong evidence that carbamazepine exists as two stable structures with different geometries (Figure 4).

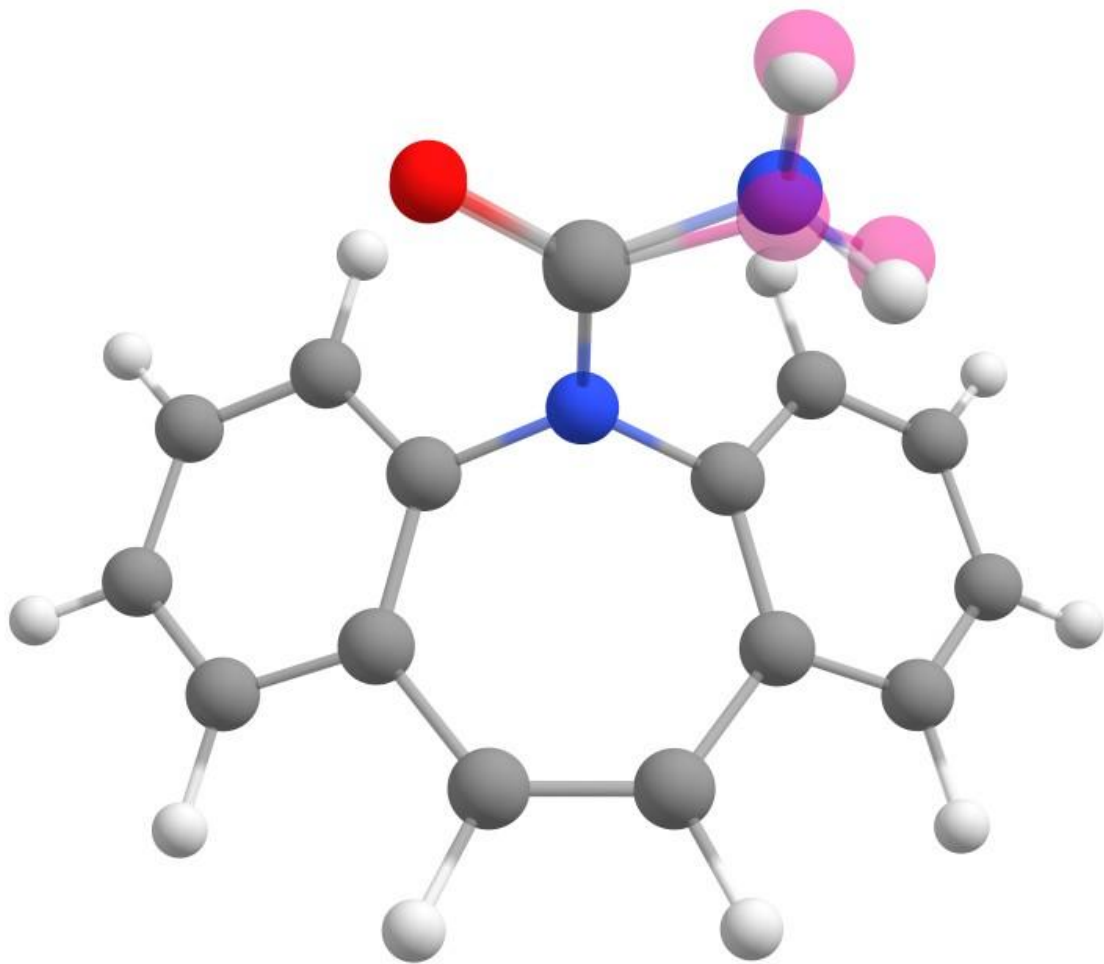

Figure 4. Comparison of carbamazipine structures of conformation 1(without atoms colored) and conformation 2 (pink). The RMSD for the two low energy conformers is 0.2 Å.

These structures exist also in the solution, which is proved by the cross-peaks in the $^1H$-$^{15}N$ HMBC spectra obtained in chloroform. Investigations performed in the $^1H$ NMR showed only one signal, which is the result of the fast proton exchange within the $NH_2$ group (Figure 5). However, analysis of the HMBC projections allows distinguishing these conformers and obtaining their relative fractions through the quantitative integration. Thus, ratio of the carbamazepine conformer fractions is 90:10, which is in good agreement with results obtained for carbamazepine in $CO_2$.

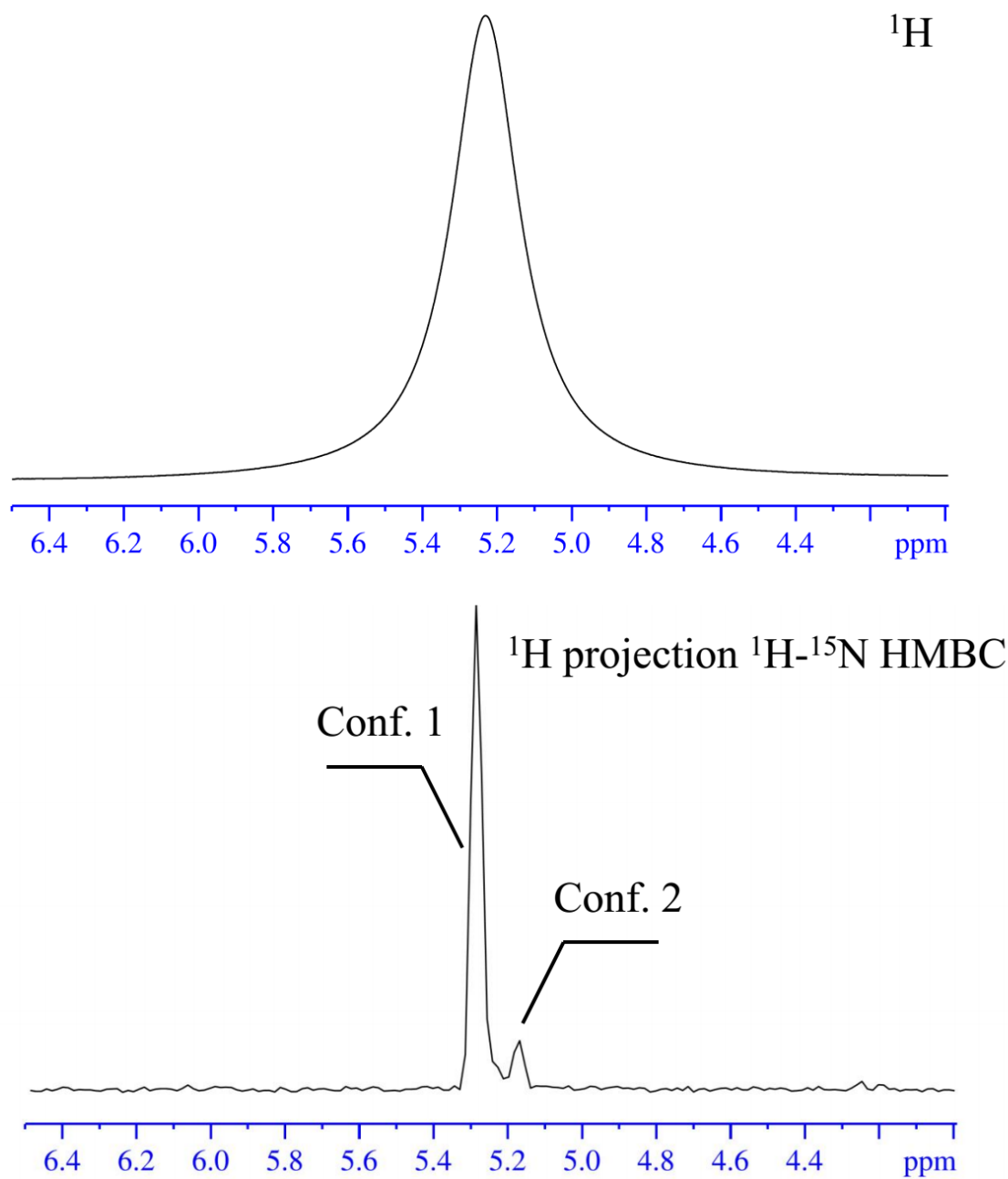

Figure 5. The $^1H$ NMR fragment of the spectrum of carbamazipin (top) and the projection through spectrum $^1H$-$^{15}N$ HMBC (bottom). The ratio of two carbamazepine conformers 1 and 2 from HMBC is 90:10, respectively.

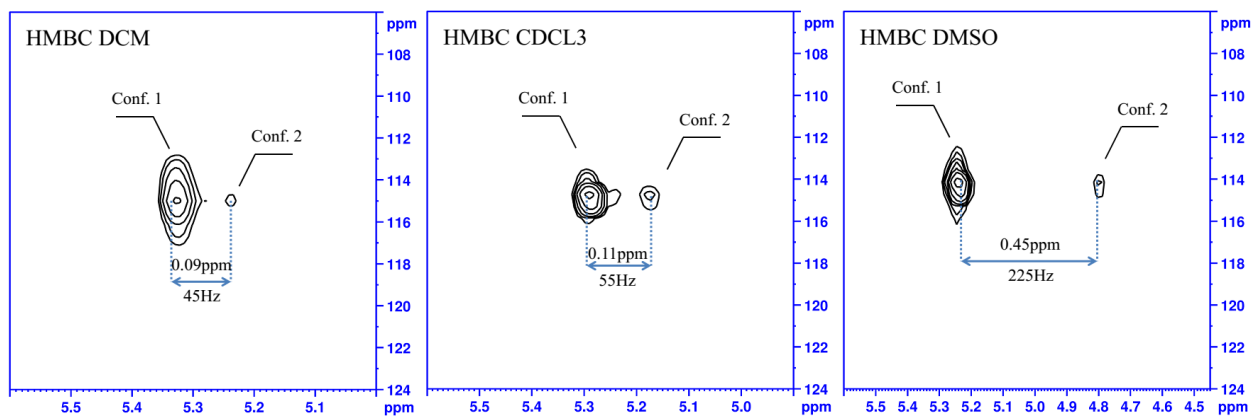

Figure 6. $^1H,^{15}N$-HMBC spectra of carbamazepine in DMC, CDCl$_3$, and DMSO. Two signals due to the presence of different conformers are observed in all considered solvents.

In addition to low-polar chloroform, spectra of saturated solutions of carbamazepine in apolar dichloromethane (DMC) and polar dimethyl sulfoxide (DMSO) were obtained. It is evident from Figure 6, that two signals appear in all three solutions. The chemical shift difference between them, however, varies from 0.10 to 0.45 ppm, which should be caused by solvent effects such as presence or absence of hydrogen bonds, magnetic anisotropy of solvent molecules, van der Waals interaction between the solvent and the solute, and polar effect due to the influence of the electric field of the solute molecules acting on the solvent molecules. In general, solvent effects were first described in 1960s by Buckingham [26] and later studied in detail by Abraham [27].

Influence of solvents onto amides is not limited to chemical shift modification. A well-known phenomena if the influence of solvents on the ratio of *cis*- and *trans*-conformers arising due to rotation around to amide bond [28]. Polar solvents can increase the energy barrier between the conformers as compared to apolar media [27]; moreover, the conformer equilibrium also often varies in CDCl$_3$ and DMSO. For instance, the *trans/cis* equilibrium varies from 1% to 3% in N-methylacetamide and from 6% to 13% in dimethyl formamide; the *exo/endo* equilibrium in N-methylformanilide at the same time varies between 5% and 11%. The conformers equilibrium of carbamazepine changes from 3% to 6% upon transition from apolar CDCl$_3$ and DMC to polar DMSO. At the same time, upon transition from CDCl$_3$ to DMC the equilibrium shifts by only 3%. The relative conformer 1 and 2 fractions of carbamazepine in DMSO is 96:4, while in DMC it is 93:7. Thus, the trends observed here agree well with the trends known for N-methylacetamide, N-methylformanilide, and N-dimethylformamide.

For instance, it would be interesting to look for the occurrence of the hidden conformers in a number of other drug compounds: sulfonamides (sulfanilamide, sulfisoxazole, sulfadiazine, sulfamethizole, sulfapyridine, sulfamethoxazole, sulfadoxine), and also other compounds: lisdexamfetamine, bethanechol, frovatriptan, in which conformation lability is induced by the mobility of the amine group.